%% file: arxiv_main_v1.tex
\documentclass[10pt,conference]{IEEEtran}
\IEEEoverridecommandlockouts
\usepackage{cite}
\usepackage{bold-extra} 
\usepackage{url}
\usepackage{setspace} 
\usepackage{multirow} 
\usepackage{amsmath,amssymb,amsfonts}
\usepackage[linesnumbered,ruled,vlined]{algorithm2e}
\usepackage{algpseudocode}
\usepackage{graphicx}
\usepackage{physics}
\usepackage{textcomp}
\usepackage{hyperref}
\usepackage{xcolor} 
\usepackage{booktabs}
\usepackage{enumitem}
\usepackage{calc}
\usepackage[draft, commentmarkup=todo,
   todonotes={textsize=tiny, textwidth=0.83in}]{changes}
\definechangesauthor[name={Bao Bach}, color=teal]{bb}
\definechangesauthor[name={Bao Bach}, color=magenta]{fbm}

\def\BibTeX{{\rm B\kern-.05em{\sc i\kern-.025em b}\kern-.08em
    T\kern-.1667em\lower.7ex\hbox{E}\kern-.125emX}}
\begin{document}
\SetKwInput{KwInput}{Input}                
\SetKwInput{KwOutput}{Output}              
\title{Solving Large-Scale QUBO with Transferred Parameters from Multilevel QAOA of low depth}

\author{
\IEEEauthorblockN{Bao G Bach}
\IEEEauthorblockA{\textit{Computer and Information Sciences} \\
\textit{Quantum Science and Engineering} \\
\textit{University of Delaware}\\
Newark DE, USA \\
baobach@udel.edu}
\and
\IEEEauthorblockN{Filip B. Maciejewski}
\IEEEauthorblockA{
\textit{Research Institute for Advanced} 
\\
\textit{Computer Science (RIACS),}
\\
\textit{Universities Space Research Association,}
\\
Moffett Field CA, USA
\\
fmaciejewski@usra.edu}
\and
\IEEEauthorblockN{Ilya Safro}
\IEEEauthorblockA{\textit{Computer and Information Sciences} \\
\textit{Physics and Astronomy} \\
\textit{University of Delaware}\\
Newark DE, USA \\
isafro@udel.edu}
}
\maketitle

\begin{abstract}
The Quantum Approximate Optimization Algorithm (QAOA) is a promising quantum approach for tackling combinatorial optimization problems. However, hardware constraints such as limited scaling and susceptibility to noise pose significant challenges when applying QAOA to large instances. To overcome these limitations, scalable hybrid multilevel strategies have been proposed. In this work we propose a fast hybrid multilevel algorithm with QAOA parameterization throughout the multilevel hierarchy and its reinforcement with genetic algorithms, which results in a high-quality, low-depth QAOA solver. Notably, we propose parameter transfer from the coarsest level to the finer level, showing that the relaxation-based coarsening preserves the problem structural information needed for QAOA parametrization. Our strategy improves the coarsening phase and leverages both Quantum Relax \& Round and genetic algorithms to incorporate $p=1$ QAOA samples effectively. 
The results highlight the practical potential of multilevel QAOA as a scalable method for combinatorial optimization on near-term quantum devices.
\end{abstract}


\input{sections/1introduction}

\input{sections/2prelim}
\input{sections/3related_works}
\input{sections/4method}
\input{sections/5experiments}
\input{sections/6conclusion}

\section*{Acknowledgment}

B.B. and I.S. are supported with funding from the Defense Advanced Research Projects Agency (DARPA) under the ONISQ program.
F.B.M. acknowledges support under the NSF awards \#2329097 and \#1918549.
We are grateful to Davide Venturelli, Joao Prioli, and José Carlos Hernández Azucena for useful discussions throughout this project.

\bibliographystyle{unsrt}
\bibliography{biblo}

\end{document}

%% file: sections/1introduction.tex
\section{INTRODUCTION}
Despite the breadth of classical algorithmic approaches, many fundamental combinatorial optimization problems remain intractable, allowing only heuristics with a reasonable time-quality trade-off to (sub)optimally solve them in practice. The class of Quantum Approximate Optimization Algorithms (QAOA) \cite{farhi2014quantum} offers a potentially transformative route to solving these challenges by harnessing quantum effects to explore solution spaces potentially more efficiently than classical methods. However, large-scale implementation of QAOA remains impractical \cite{guerreschi2019qaoa, maciejewski2024design} due to hardware constraints such as limited connectivity, short coherence times, and high error rates (not to mention the number of qubits), which significantly restrict the size and depth of circuits that can be reliably executed \cite{zhou2020limits}. These hardware limitations present significant obstacles to realizing the full potential of QAOA in real-world applications. 

A promising strategy to mitigate these limitations is the multilevel approach \cite{ushijima2021multilevel}, inspired and generalized by algebraic multigrid techniques \cite{brandt2003multigrid}. The guiding principle of the method is summarized by the motto, “Think globally but act locally.” A large optimization problem is repeatedly coarsened to create smaller, more manageable instances, each of which can be decomposed into parts or tackled as a whole using QAOA with fewer qubits. At each coarse level 
$i$, the best-found solution becomes the initialization for the next finer level $i-1$, preserving target optimization information across scales. This initialization undergoes ``local processing" (or refinement), a scalable and effective procedure that modifies only a small subset of variables at a time but systematically revisits all variables at that level. The refined result is progressively uncoarsened and refined, ultimately mapping back to the original, fine-scale formulation. While multilevel QAOA \cite{ushijima2021multilevel,bach2024mlqaoa ,maciejewski2024multilevel} has shown promise in accommodating hardware restrictions, existing multilevel QAOA implementations can be restrictive in two key respects: (1) the variational parameters are not inherited throughout uncoarsening, and (2) QAOA is not hybridized with fast classical approaches to fully exploit joint potential at all levels of coarseness.

\noindent {\bf Our contribution} To address these issues, we propose an enhanced multilevel QAOA framework that incorporates a novel parameter transfer mechanism. The parameters identified at coarser levels are carried forward to finer levels, highly reducing the need for parameter optimization (and thus the required quantum resources) -- see Fig.~\ref{fig:param-transfer} for illustration. This is because the relaxation-based coarsening preserves essential QUBO information—a claim we support with numerical evidence. This method capitalizes on the insight that single-layer QAOA solutions \cite{ozaeta2022expectation} on coarse graphs can capture core structural features, thereby providing a strong starting point for finer-level optimization. 

Alongside improved coarsening phases, our framework fuses the Quantum Relax \& Round \cite{dupont2024extending} and genetic algorithms 
to effectively incorporate single-layer QAOA samples. Through comprehensive numerical experiments, we show that this combination of careful coarsening, guided parameter transfer, and iterative refinement boosts the solution quality and also maintains a high level of efficiency. Overall, our work demonstrates the potential of multilevel QAOA as a powerful, scalable strategy for tackling real-world combinatorial optimization problems on near-term quantum devices even with low depth circuits.

%% file: sections/2prelim.tex
\section{PRELIMINARIES}
\subsection{Quantum Approximate Optimization Algorithm}
The Quantum approximate optimization algorithm (QAOA) \cite{farhi2014quantum} is a hybrid quantum algorithm proposed to tackle combinatorial optimization problems in the form of the Ising Model.
Given the Ising model, QAOA with $p$ layers alternatingly apply unitaries drawn from two Hamiltonian families, cost unitary $U_{P}(\gamma) = e^{-i\gamma H_f}$ and mixing unitary $U_{M}(\beta) = e^{-i\beta H_B}$ parametrized by $\gamma = \{\gamma_i\}$ and $\beta = \{\beta_i\}$, $1\leq i \leq p$, respectively. Hamiltonian $H_f$ is an Ising cost Hamiltonian where the information of the given Ising model is embedded, while Hamiltonian $H_B$ is a fixed mixing Hamiltonian. Using $H_f$ as the observable, QAOA prepares the quantum state expressed in (\ref{eq: QAOA_state}) and performs optimization concerning the expectation value $\langle H_f \rangle = \bra{\gamma,\beta}H_f\ket{\gamma,\beta}$ to find the minimum.
\begin{equation}
    \label{eq: QAOA_state}
    \ket{\gamma,\beta} = U_{M}(\beta_p)U_{P}(\gamma_p)\dots U_{M}(\beta_1)U_{P}(\gamma_1)\ket{+}^{\otimes n}
\end{equation}

\subsection{Problem Formulations}
Here, we define an $n-$variable Quadratic Unconstrained Binary Optimization (QUBO) problem as a \textit{maximization} problem over a binary vector $x \in \mathbb{F}^{n}_{2}$. The QUBO problem is specified by real-valued, upper-triangular matrix $Q \in \mathbb{R}^{n\times n}$ where entries $Q_{i,j}$ denotes weight for each pair of indices $i, j\in \{0,\dots, n-1\}$. QUBO has the cost function of $\max_{x}  \sum^{n-1}_{i=0}\sum^{n-1}_{j=i}Q_{i,j}x_{i}x_{j}$.

Among QUBO problems, we are interested in the MAXCUT problem, which is an NP-complete problem 
and is commonly used as the benchmark for testing QUBO \cite{dunning2018works}. Solving MAXCUT problem w.r.t $n-$node graph $G(V, E)$ with $\abs{V} = n$ involves finding a cut that splits the graph node into two disjoint partition, $V_1$ and $V_2$, such that the weighted sum of edges $ij$ connecting the two parts is maximized. Graph $G(V, E)$ can be represented as upper-triangular, zero diagonal  matrix $W \in \mathbb{R}^{n\times n}$ and having the \textit{maximization} cost function of $\max_{x} \sum^{n-1}_{i=0}\sum^{n-1}_{j=i}W_{i,j}(x_i + x_j - 2x_ix_j )$. Given $ n-$variable QUBO problem, it can be transformed to $n+1$-node MAXCUT problem up to a multiplicative factor by mapping $W_{i,j} = Q_{i, j}$ for $i \neq j$ and $W_{i, n} = Q_{i, i} + \sum^{n-1}_{j=i}Q_{i, j}$ for $i < n$ \cite{dunning2018works}. Given the solution of MAXCUT problem denoted as $y$, the solution of the original QUBO problem can be retrieved via $x_{i} = y_{i} \bigoplus_{2}y_{n}$, where $\bigoplus_{2}$ is addition modulo 2. As the MAXCUT problem can be equivalently formulated as the Ising model without external fields 
, to solve $n-$variable QUBO problem using QAOA, we first reformulate it as $n+1$-node MAXCUT and identify it with $n+1$-spin Ising Hamiltonian. 

\subsection{Quantum Relax \& Round}
Optimizing an $n-$spin Ising Hamiltonian using QAOA yields a final quantum state $\ket{\psi^{*}} = \ket{\gamma^{*}, \beta^{*}}$ where $\gamma^{*}$ and $\beta^{*}$ are the optimal parameters. Building on this state, \cite{dupont2024extending} proposed an efficient Quantum Relax \& Round (QRR) algorithm inspired by classical relax \& round strategies (with extensions presented in \cite{dupont2025optimization}).
By approximating the two-point correlations of spins denoted by $Z \in \mathbb{R}^{n \times n}$ with entries $Z_{i,j}= (\delta_{ij} - 1)\langle\sigma^{(i)}_{z}\sigma^{(j)}_{z}\rangle_{\gamma, \beta}$ where $\delta_{ij}$ is the Kronecker delta, QRR performs an eigendecomposition of $Z$, sign-rounds the resulting eigenvectors, and evaluates these rounded vectors as potential solutions.

Due to the limited number of qubits available on current quantum hardware \cite{horowitz2019quantum} and the difficulty of simulating larger quantum systems \cite{zhou2020limits}, collecting extensive samples for large-scale problems remains infeasible. To overcome this constraint, we employ QRR specifically for the $p=1$ QAOA regime, where analytic expressions for spin correlations are available \cite{ozaeta2022expectation}. This approach enables us to form the correlation matrix $Z$ directly from analytically computed values and apply QRR without requiring extensive sampling, thereby generating candidate bit strings for larger problem instances despite hardware and simulation limitations.

\subsection{Multilevel Method}

The multilevel method (also known as multiscale, multiresolution, and multigrid-inspired methods)\cite{brandt2003multigrid} is widely employed to tackle large-scale computational problems on different hardware architectures. Especially, the multilevel method has been explored in the context of quantum hardware with different problems, such as linear arrangement, graph partitioning, and MAXCUT with QAOA \cite{ushijima2021multilevel, angone2023hybrid, bach2024mlqaoa, maciejewski2024multilevel}. Guided by the motto, ``Think globally but act locally", the multilevel method iteratively transforms the original problem to its coarser version. When dealing with QUBO problems represented by graphs, this process typically merges pairs of nodes according to an algebraic distance metric \cite{chen2011algebraic}, where the choice of merging aims to preserve problem-relevant structure \cite{angone2023hybrid, bach2024mlqaoa}. Once a solution is obtained for a coarser graph at level $i$, it serves as the initialization point for refinement at the next finer level $i-1$, repeatedly “uncoarsening” until the original, fine-scale problem is recovered. In other words, solutions flow from coarse to fine levels, reducing complexity at each step while retaining global insight.

While there are multiple schemes for these coarsening-uncoarsening phases, as proposed in the multigrid literature \cite{brandt2003multigrid}, in this study, we adopt the canonical, simplest  V-cycle scheme, explicitly to observe the effects of $p=1$ QAOA in a controlled setting, and to minimize additional advantages from advanced classical preprocessing. Concretely, we define a hierarchy of graphs as $\mathcal{L} = \{G_{l}(V_{l}, E_{l}, w_{l})^{L}_{l=0} \}$. Here, graph $G_{l}$ contains a set of vertices $V_{l}$, a set of edges $E_{l}$, and a set of edge weights $w_{l}$ and denotes the transformed graph at level $l$. Starting from the coarsest level $G_{L}$, we successively uncoarsen and refine the solution, reintroducing details at each finer level until reaching the finest level $G_0$ (original graph). This systematic, locally driven refinement procedure aligns well with quantum hardware constraints, which limit the number of qubits but can still deliver valuable partial solutions that, when interpolated, produce strong overall performance for large-scale optimization problems.

%% file: sections/3related_works.tex
\section{Related Work}
\subsection{QAOA for large graphs}
Quantum Approximate Optimization\cite{farhi2014quantum} is a promising quantum algorithm candidate to tackle combinatorial optimizations \cite{shaydulin2024evidence, farhi2016quantum}. However, the current noisy and intermediate-scale quantum hardware creates a significant hindrance in implementing this algorithm \cite{zhou2020limits, maciejewski2024design, maciejewski2024improving, hao2024end}. Consequently, when considering large-scale optimization problems, this matter worsens as the quantum hardware's limitations prevent implementing QAOA for the whole problem. To overcome these challenges, researchers have been looking into various strategies, including
developing qubit-efficient encodings to enable the use of QAOA for large-scale problems \cite{sciorilli2025towards, fuller2024approximate, sundar2024qubit} or performing classical preprocessing to downscale the original problems into manageable subproblems \cite{ushijima2021multilevel, angone2023hybrid, bach2024mlqaoa, ponce2025graph,acharya2024decomposition}.

In addition, researchers are looking into utilizing distributed quantum computers to reduce the scaling problem. A candidate for this is using divide-and-conquer methods to tackle divided subgraphs in parallel and then merging these solutions to derive the final solution \cite{li2022large, tomesh2023divide, cameron2024scaling}. From a quantum hardware perspective, approaches such as circuit cutting \cite{bechtold2023investigating, perlin2021quantum, dupont2025benchmarking} are used for cutting big quantum circuits to allow running on smaller quantum processing units (QPU).

\subsection{Parameter setting for hybrid quantum-classical approach}
Given a parameterized quantum circuit, finding the optimal parameters to minimize a given observable is a hard task and, in some cases, NP-hard \cite{bittel2021training}. To make it worse, recent research observes that the landscape of variational quantum algorithms often exhibits barren plateaus when the circuit depth is large for big-scale problems \cite{larocca2024review}, which might also apply to QAOA \cite{rajakumar2024trainability}. Furthermore, by increasing the number of layers to improve the approximation result from QAOA, the computational complexity to optimize and evaluate the parameterized circuit increases. Therefore, optimal parameter search acceleration strategies have been explored with multistart \cite{shaydulin2019multistart} or warmstart \cite{egger2021warm} methods. In parameter transfer methods, the potentially suboptimal performance is traded for avoidance of the costly optimization steps through the use of a fixed set of parameters for multiple problem instances, irrespective of problem type \cite{shaydulin2023parameter, galda2023similarity, nguyen2025cross}. 

%% file: sections/4method.tex
\section{Methods}
The Multilevel QAOA framework \cite{angone2023hybrid, bach2024mlqaoa, maciejewski2024multilevel} provides a way to use small quantum devices to facilitate solving large instances of QUBO\footnote{Note that we will mention QUBO and MAXCUT interchangeably as describing coarsening is easier with graph. Our experimental instances are both positive and negative weighted on $w_{ij}$ which is beyond the traditional MAXCUT formulation.}.
However, it involves certain challenges, such as (1) the coarsening scheme can mistakenly merge nodes that are irrelevant due to the pairwise constraint, and (2)  the suboptimal sub-problem graph selection can lead to no improvement during the refinement process.
Moreover, to efficiently tackle these sub-problem graphs, a carefully designed acceleration scheme \cite{bach2024mlqaoa} has to be devised to avoid the computational overhead caused by the variational loops.
To address these limitations, we modify the algorithm to relax the coarsening phase constraints in section \ref{sec:coarsening}. Moreover, for QAOA-based optimization, we numerically observe that for $p=1$ circuits, the optimal parameters are approximately transferred between the circuits at \textit{different} levels of the  hierarchy, effectively reducing the need for parameter optimization for all levels except the coarsest.

\subsection{Coarsening Phase}
\label{sec:coarsening}
The target of this process is to create a coarser version of the current graph so that the information related to the target problem, such as QUBO, is preserved.
Based on the well-known strategy of the Goemans-Williamson method for solving MAXCUT 
by formulating the problem into a Semidefinite Programming (SDP) framework, the coarsening starts with randomly embedding the nodes of graph $G_l$ onto a surface of a $3-$dimensional sphere. 
Given $p^{t}_i$ as the cartesian coordinates of node $i$ at given level $t$, $\forall i\in V_{l}, \norm{p^{t}_i}^{2}_{2} = 1$. 
Following the nodes' initialization, a heuristic node-wise iterative correction is applied to maximize the weighted distance between each node and its neighbors.
\begin{align}
    \label{eq:relax_based_method}
    \forall {i} \in V_l, & \max_{p^{t}_{i}}{\sum_{j \in N(i)}} w_{ij} \norm{p^{t}_{i} - p^{t}_{j}}_{2}\\
    \textit{s.t:  } & p^{t}_{i} - 1 = 0 \nonumber
\end{align}
where $N(i)$ is set of neighbors of node $i$, and $w_{ij}$ is weight of edge $ij$. Here, our heuristic follows the strategy of Weiszfeld's algorithm 
with assumption $\norm{p^{t}_{i} - p^{t}_{j}}_{2} \approx 1$ and achieves a closed-form solution for each point (i.e., node) $i \in V_{l}$ through the Lagrange‑multiplier solution of Equation \ref{eq:relax_based_method}. 
The iterative scheme continues until convergence.

\begin{figure}
    \centering
    \includegraphics[width=\linewidth]{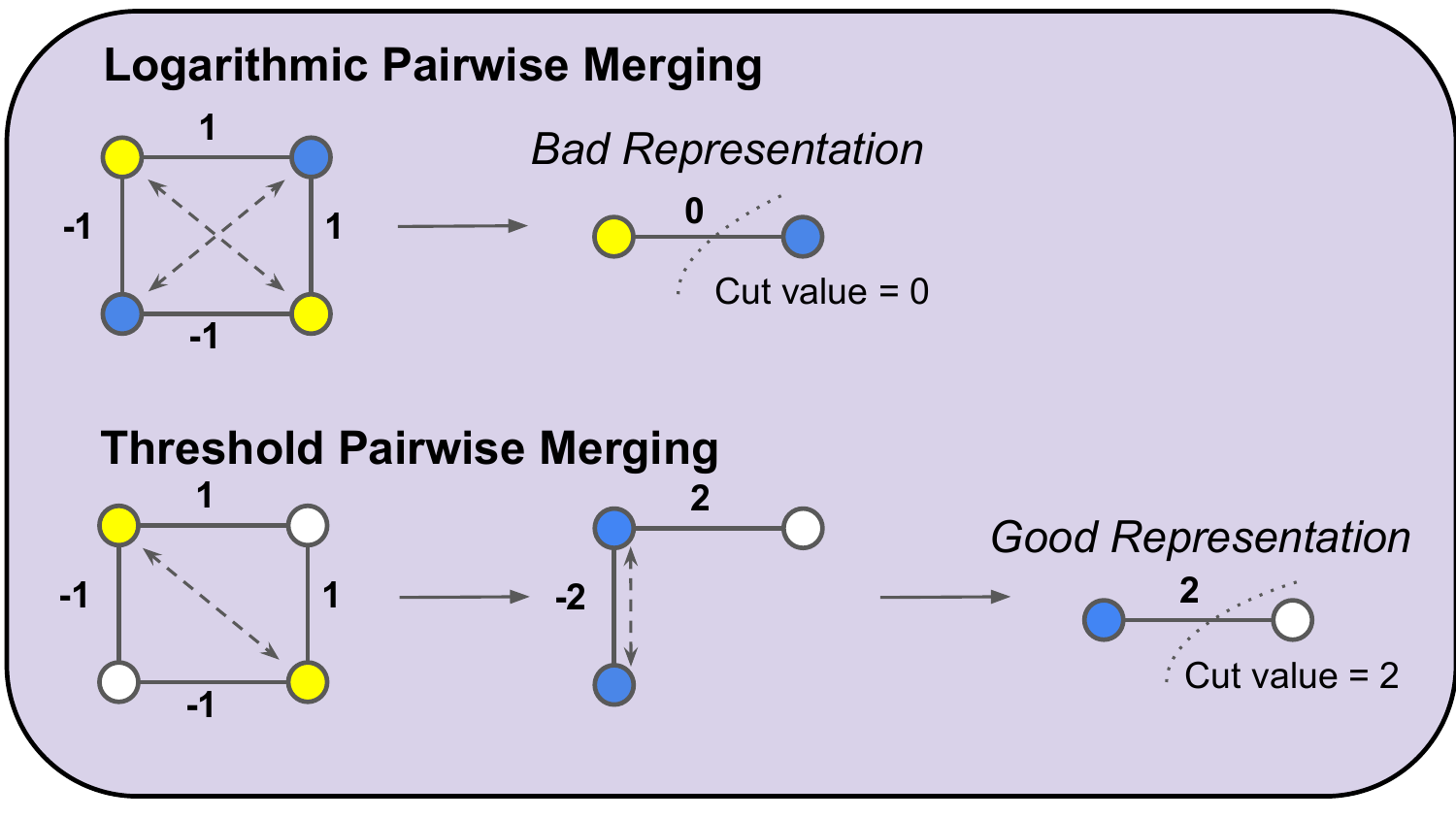}
    \caption{Example between two merging strategies: (1) Logarithmic Pairwise Merging (forcefully merge pairs of nodes so that the size of the next-level graph is half the current level), and (2) Threshold Pairwise Merging (merge pairs of nodes only if the distance between them is low enough). Here, the first strategy fails to preserve the QUBO information of the original graph. The second strategy introduces one more level but captures the original graph cost correctly.}
    \label{fig:coarsening_intuition}
\end{figure}

After embedding, the current configuration of points on the sphere encodes information about the target problem by separating nodes from their neighbors connected by positive weight while bounding neighbors connected by positive weight. 
From this, we use the K-D tree \cite{friedman1977algorithm} to compute the pairwise distance between nodes and their nearest neighbors on the sphere. 
It is important to note that in the previous approaches ~\cite{bach2024mlqaoa,maciejewski2024multilevel} (referred as the Logarithmic Pairwise Merging), one would merge $\frac{|V_l|}{2}$ pairs according to the smallest pairwise distances to ensure that the total number of coarsened graphs scales logarithmically with the problem size, thus reducing the overall algorithmic complexity. 
However, this approach can sometimes lead to merging nodes with relatively high distances, converging to lower quality coarsened representations. 
Here, we propose a more refined approach, a Threshold Pairwise Merging, where the tradeoff between the number of coarsened graphs and the representation fidelity is controlled by threshold hyperparameters $\Delta_{1}, \Delta_{2}$, which determine how strictly we want to merge the nodes.
In particular, based on the list of pairwise distances, we start pairing nodes \textit{only if for two nodes $i$ and $j$, we have $d_{ij} \leq \max\{\Delta_1\bar{d}, \Delta_2\min d\}$} and $w_{ij} \leq 0$, where $d$ is the matrix of pairwise distances, and $\bar{d}$ ($\min d$) denotes the average (minimum) distance. 
By merging all nodes paired in the matching, we create the coarser graph. 
The intuition for enforcing this constraint comes from the fact that merging the nodes with large distances can lead to a false representation of the coarser level that fails to capture the best solution. 
In Figure~\ref{fig:coarsening_intuition}, we show an illustrative example where the Logarithmic Merging, which ignores the distance, leads to a bad representation, while enforcing the above-described constraints in Threshold Merging preserves the QUBO information correctly at the coarser level (leading, however, to one additional level).
This condition may result in more levels, at the gain of improved quality.

\label{sec:param_transfer}
\begin{figure}
    \centering
    \includegraphics[width=\linewidth]{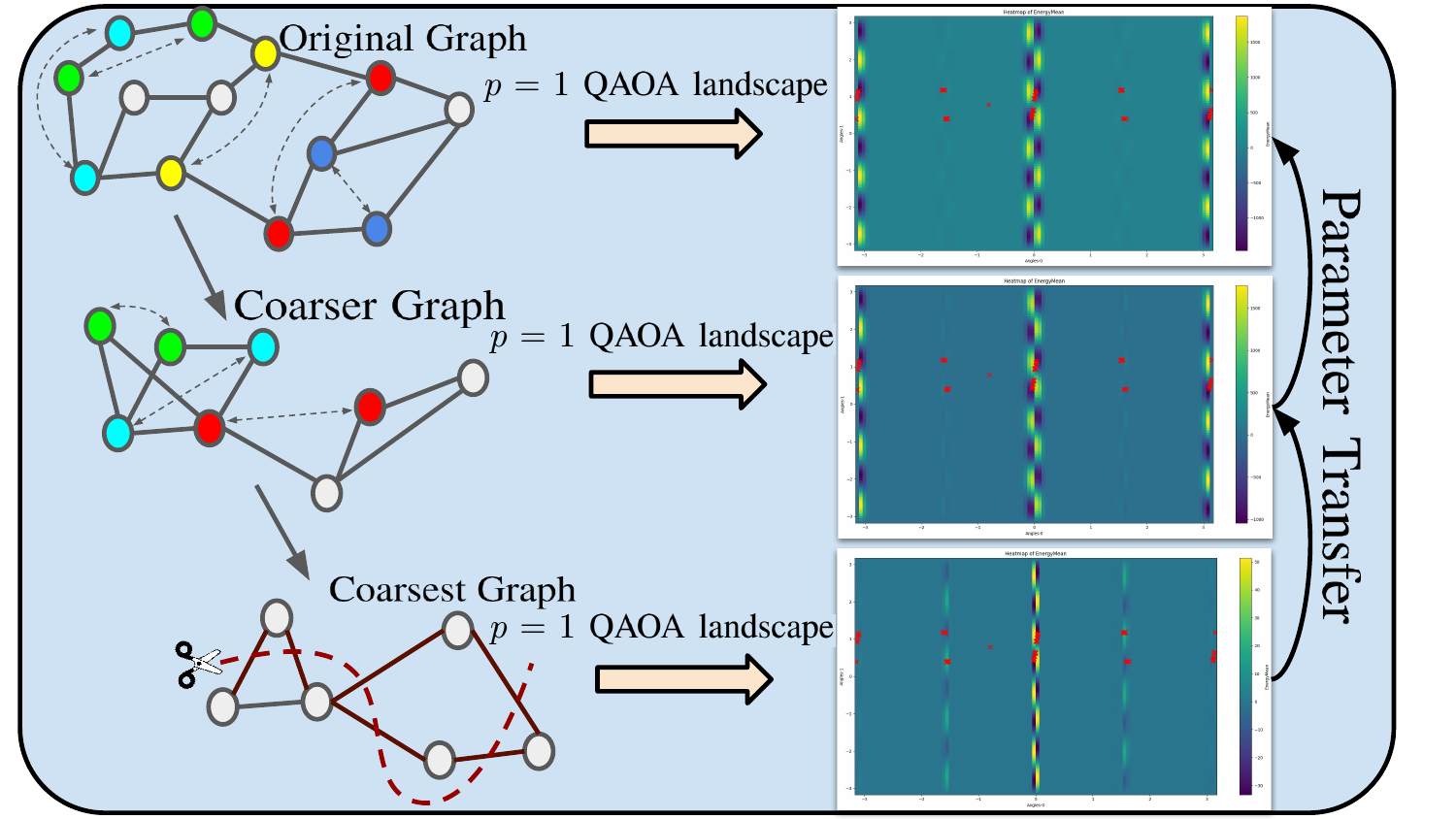}
    \caption{
    Illustration showing how the $p=1$ QAOA landscape changes with coarsening hierarchy levels.
    The empirically observed similarity between the non-trivial regions of the landscape motivates the parameter transfer between levels.
    In the plots on the right, we additionally denote the top $s$ QRR solutions (see the main text) as red points.
    }
    \label{fig:param-transfer}
\end{figure}


\subsection{Uncoarsening Phase}
\label{sec:uncoarsening}

Given the hierarchy of progressively coarser problems, a single node in a graph of a given level represents multiple nodes from finer levels, a concept central to multilevel methods.
However, this does not fully reflect the impact of quantum algorithms such as QAOA during each level. 
To explore the integration of quantum algorithms more fully while avoiding hardware limitations, we perform large-scale simulations based on the \textit{scalable} analytical form of $p=1$ QAOA \cite{ozaeta2022expectation} (note that for higher $p$, QAOA is generally hard to simulate). At the coarsest level, a simple grid optimizer identifies suitable parameter sets $\{(\gamma_i, \beta_i)\}^{k}_{i = 0}$ for which we also store the resulting set of correlation matrices $\{Z_{\gamma_i,\beta_i}\}^{k}_{i=0}$. These matrices encapsulate the two-point spin correlations under $p=1$ QAOA. We then perform Quantum Relax \& Round to each matrix $Z$ to produce the top candidate solutions for each set of parameters. The best $s$ of those candidates (across all parameter sets) are then used as an initial population of the genetic algorithm (GA), which generates and outputs the best solution at the coarsest level. Genetic algorithm is a class of search and optimization algorithms inspired by the principles of natural selection and genetics. Specifically, genetic algorithms
treat candidate solutions as ``chromosomes" and perform genetic operators such as selection, crossover, and mutation. The best solution evaluated by fitness equation is returned. 
We note that GA is implemented entirely classically on the level of measurement outcomes (bitstrings).

Each subsequent refinement phase (Algorithm \ref{algo:Refinement}) begins by linearly interpolating the coarsest-level solution to the finer graph via a surjective mapping $F: V_{l+1} \rightarrow V_{l}$. In other words, the bitstring assignment for each node $i$ at the finer level $l-1$ is initialized to the value of its corresponding node $F(i)$ at the coarser level $l$. Moreover, when moving to the next level, we reduce the QAOA parameter-grid size to include only the parameters corresponding to the top $s$ solutions from the previous level, thus exploiting the parameter transfer between levels.
These $s$ candidates, along with the interpolation of the best solution from the coarsest level, are then used as the initial population of the genetic algorithm. 

This process continues until the finest level (the original graph) is reached. 
While our current implementation uses QRR to generate bitstring candidates (since we can only simulate local expected values at large scale), the same principle could be applied directly to the sampled measurement outcomes in experimental implementation (which could then be combined with QRR-derived solutions, where one would simply choose whichever is better).
In either scenario, the refinement procedure adds minimal overhead (no optimization cost), given that only the top solution from each level needs to be updated, and quantum measurements or correlation-matrix computations are performed sparingly due to the parameter transfer. Consequently, this approach strikes a practical balance: it leverages the benefits of QAOA-based refinements at multiple levels without overwhelming hardware constraints, opening a pathway to more scalable quantum-assisted optimization.
\begin{algorithm}
    \DontPrintSemicolon
    \caption{Refinement}
    \label{algo:Refinement}
    \KwInput{Graph $G_{l}(V_{l},E_{l})$, initial solution from previous level $S_{l+1}$, good parameter sets from previous level $\{(\gamma_{i}, \beta_{i})\}^{k}_{i=0}$.}
    \KwOutput{Good parameter sets at current level $\{(\gamma_{i}', \beta_{i}')\}^{k'}_{i=0}$, best solution $S$.}
    Linearly interpolate solution $S$ for graph $G_l$.\\
    Initialize $best\_candidates \gets S$.\\
    \For{$(\gamma_{i}, \beta_{i})$ in $\{(\gamma_{i}, \beta_{i})\}^{k}_{i=0}$ }
    {
        Perform \textit{QRR} with $Z_{\gamma_{i}, \beta_{i}}$ obtained from QAOA.\\
        Obtain the best $n$ bitstring as candidate solutions.\\
        Save $m$ candidate to $best\_candidates$.\\
    }
    Select top $n$-th from $k\times m$ best candidates \\
    Initialize $\{(\gamma_{i}', \beta_{i}')\}^{k'}_{i=0}\}$  with parameters from top $n$-th candidates, and perform \textit{Genetic Algorithm} (GA) on those candidates. \\
    Obtain the best solution $S$ as output of GA.
\end{algorithm}

%% file: sections/5experiments.tex
\section{COMPUTATIONAL RESULTS}

For considered system sizes, the solutions obtained from the multilevel QAOA $p=1$ solver are not of sufficient quality to compete with state-of-the-art (SoTA) classical solvers.
However, we observe that \textit{using solutions from the ML solver as an initial guess (warm start) for the SoTA solver (Burer-Monteiro SDP \cite{burer2002rank}) can lead to solutions better than the raw Burer-Monteiro implementation with the same or higher runtime} (at least $800$s higher in practice). 
In what follows, we investigate the performance of such a hybrid solver.
We look at the Gset dataset \cite{yyyeGset} with graph sizes from $5000$ nodes up to $10000$ nodes, and graphs from the SuiteSparse Matrix Collection \cite{davis2011university} going up to $14000$ nodes.
The genetic algorithm is implemented using PyGAD ~\cite{gad2024pygad}.
Note that, in our experiments, QAOA expectation values are simulated classically, which has $O(n^2)$ complexity for $p=1$, hence it can be quite demanding for large-scale graphs.
Those costs could potentially be reduced with the use of quantum computers. Moreover, they would also facilitate more performant optimization with higher-depth QAOA circuits.
Our source code and results are available at [link will be provided upon acceptance]. 

\subsection{$p=1$ QAOA landscape analysis}
\label{sec:landscape_analysis}
\begin{figure}
    \centering
    \includegraphics[width=.86\linewidth]{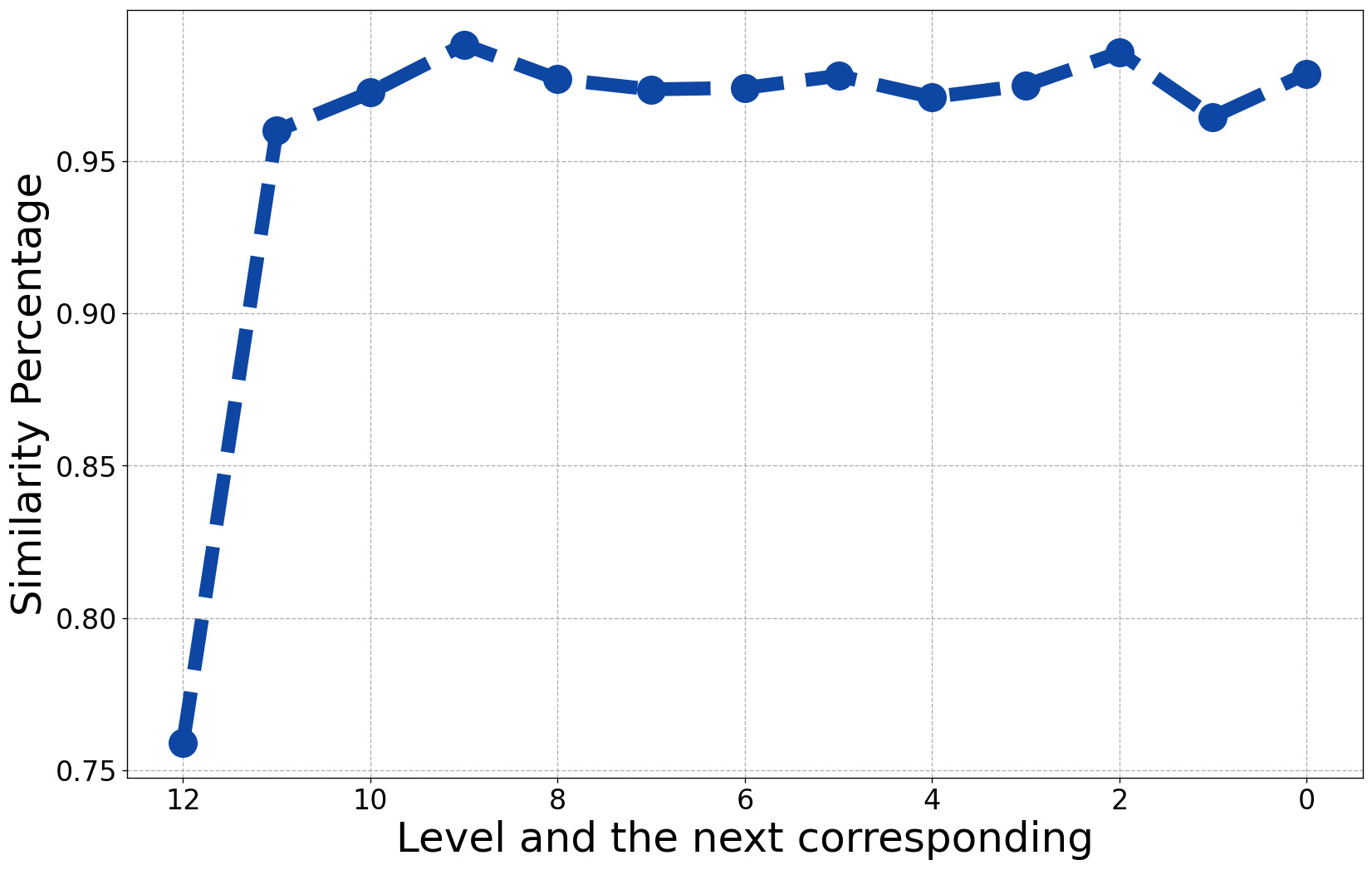}
    \caption{An example of landscape similarity between consecutive hierarchy levels. The similarity is calculated between levels $i$ and $i+1$ using the figure of merit described in the main text. Level $0$ denotes the finest graph, whereas level $12$ denotes the coarsest graph.     
    }
    \label{fig:similarity_percentage}
\end{figure}

We start by analyzing the effectiveness of the parameter transfer between levels for $p=1$ QAOA landscape.
To achieve this, we use the following analysis on the grid-based landscape where each point represents a parameter $(\gamma,\beta)$. First, to capture all the ``interesting" zones (regions in neighborhoods of local extrema), we detect all the grid points for which $\langle H\rangle_{(\gamma, \beta)} > 3\sigma_{(\gamma,\beta)}$, where $\sigma_{\gamma,\beta}$ denotes the standard deviation of average energy w.r.t. variational paramters.
Thus, we look at all of the regions that are far from the average value. 
Then, for given two hierarchy levels, we take the detected zones, and calculate the size (i.e., the number of grid points) of their intersection, normalized by the size of their union, which we treat as a heuristic similarity measure between two landscapes. Figure \ref{fig:similarity_percentage} shows a concrete example of how the similarity between landscapes in consecutive hierarchy levels changes.
We observe that for most hierarchy levels, the QAOA energy landscapes exhibit a high degree of similarity ($\gtrsim 95\%$).
The anomalous point with $\sim 75\%$ similarity corresponds to the coarsest level, where a large reduction in the graph size caused the landscape to change more significantly.
Despite being smaller than for the rest of the levels, it still constitutes a fairly high level of similarity; thus, when performing the parameter transfer and pruning scheme, the optimized parameters at the coarsest level are expected to perform well in further levels.

\subsection{Solver performance}
\begin{table*}[h]
\centering
\caption{Results of our method over different graphs with different graph sizes. Here, we compare our method with well-known classical approaches, and the best objective is recorded. The best results are in bold. The corresponding references for (much slower than multilevel) classical global solvers are BURER02 \cite{burer2002rank}, FES02GVP \cite{festa2002randomized}, DUARTE05 \cite{DUARTE2005}, and GLOVER10 \cite{glover2010diversification}.}
\label{tab:exp_res}
\small 
\begin{tabular}{l|l|l|l|l|l|l|l|l|l}
\hline
\multicolumn{4}{c|}{} & \multicolumn{2}{|c|}{\footnotesize \textbf{Our Multilevel Approach}} & \multicolumn{3}{|c}{\footnotesize \textbf{Global Classical Approach}} \\ \hline
\footnotesize \textbf{Graph} & \footnotesize $\abs{V}$ & \footnotesize $\abs{E}$ & \footnotesize $w_{E}$ & \footnotesize QAOA+BURER02 & \footnotesize ML\_BURER02 & \footnotesize BURER02  & \footnotesize FES02GVP  & \footnotesize DUARTE05  & \footnotesize GLOVER10  \\ \hline
$facebook$ & 3437   & 49138 & \{1\}     & 28533    & \bf{28559} &  28553  &  28498 & 28545 & 28326  \\
$G_{55}$   & 5000   & 12498 & \{1\}     & \bf{10273} & 10250 & 10280  &  10150 &  10155 & 10043 \\ 
$G_{56}$   & 5000   & 12498 & \{-1, 1\} & 3986  & \bf{3990}  & 3984  &  3882  &  3842 & 3779\\ 
$G_{57}$   & 5000   & 10000 & \{-1, 1\} & 3446  & \bf{3452} & 3444  &  3438 &  3382 & 3134\\ 
$G_{58}$   & 5000   & 29570 & \{1\}     & 19164 & \bf{19171} & 19131 & 19099 & 19090 & 18970\\ 
$G_{59}$   & 5000   & 29570 & \{-1, 1\} & 5965  & \bf{5969} & 5958  & 5888 &  5893 &  5808\\ 
$G_{60}$   & 7000   & 17148 & \{1\}     & \bf{14146} & 14125 & 14141   & 13991 &  13935  & 13851\\ 
$G_{61}$   & 7000   & 17148 & \{-1, 1\} & 5741       & 5731 & \bf{5745}   & 5579 &  5534 &  5446\\ 
$G_{62}$   & 7000   & 14000 & \{-1, 1\} & 4790       & 4780 & \bf{4796}   & 4764 &  4708  & 4356\\
$G_{63}$   & 7000   & 41459 & \{1\}     & \bf{26838} & \bf{26838} & 26809   & 26717 & 26744 & 26611\\
$G_{64}$   & 7000   & 41459 & \{-1, 1\} & \bf{8580}  & 8565 & 8573   & 8452 &  8448 & 8376\\ 
$G_{65}$   & 8000   & 16000 & \{-1, 1\} & 5460       & \bf{5478} & 5476   & 5416 & 5378 & 4960 \\ 
$G_{66}$   & 9000   & 18000 & \{-1, 1\} & 6240       & 6220 &\bf{6242}   & 6142 & 6150 & 5694 \\ 
$G_{67}$   & 10000  & 20000 & \{-1, 1\} & \bf{6830}  & 6826 &\bf{6830}   & 6726 &  6720 &  6144 \\ 
$G_{70}$   & 10000  & 9999  & \{1\}     & \bf{9533}  & 9532 & 9527  & 9442 &  9310 &  9136 \\ 
$G_{72}$   & 10000  & 20000 & \{-1, 1\} & \bf{6882}  & 6866 & 6868   & 6762 & 6768 & 6192 \\ 
$rajat06$  & 10923  & 18061 & \{1\}     & {\bf 14431}& 14367 &14418  & 14407  & 14211 & 13203 \\
\hline
\end{tabular}
\end{table*}

\begin{figure}
    \centering
    \includegraphics[width=.88\linewidth]{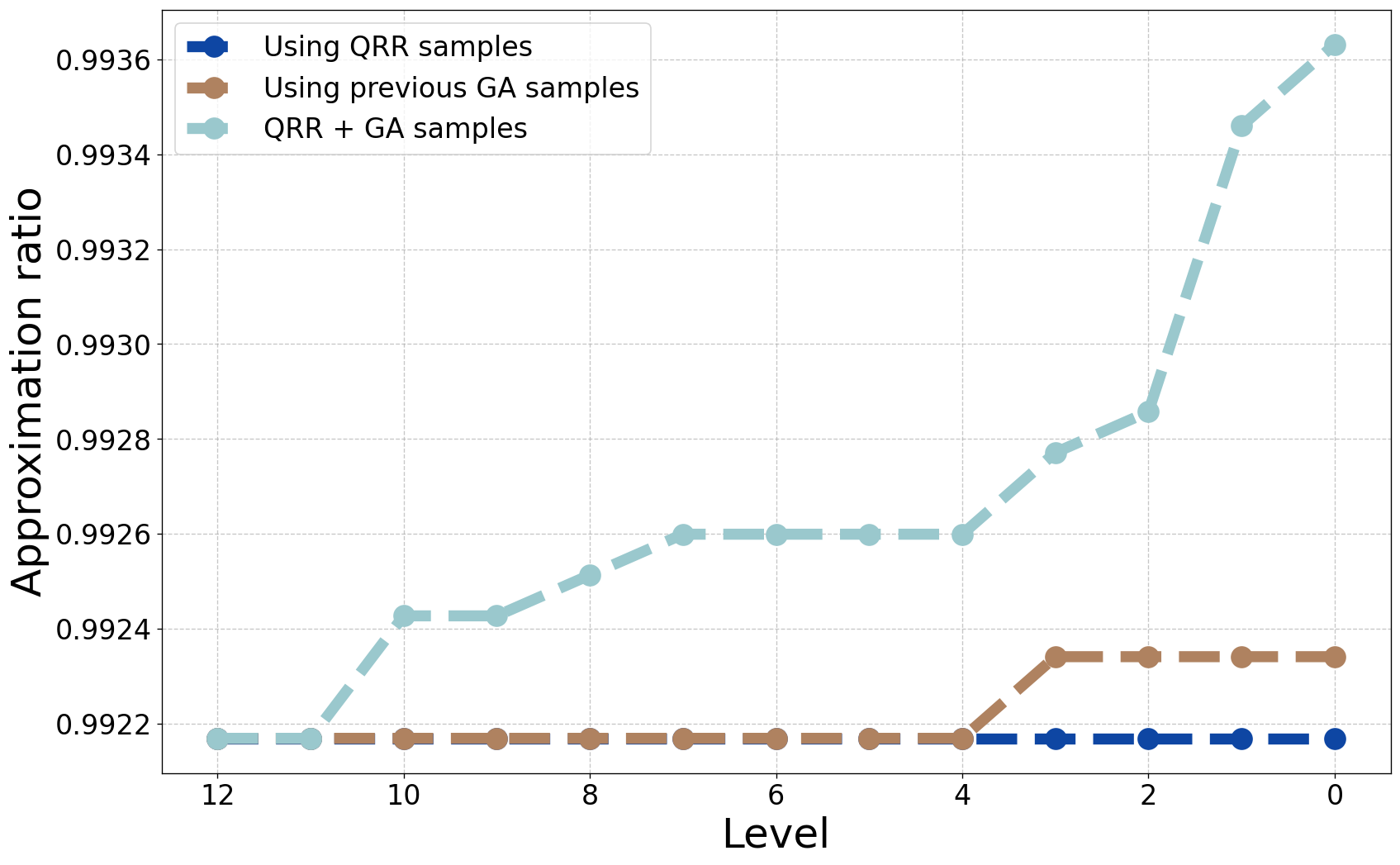}
    \includegraphics[width=.88\linewidth]{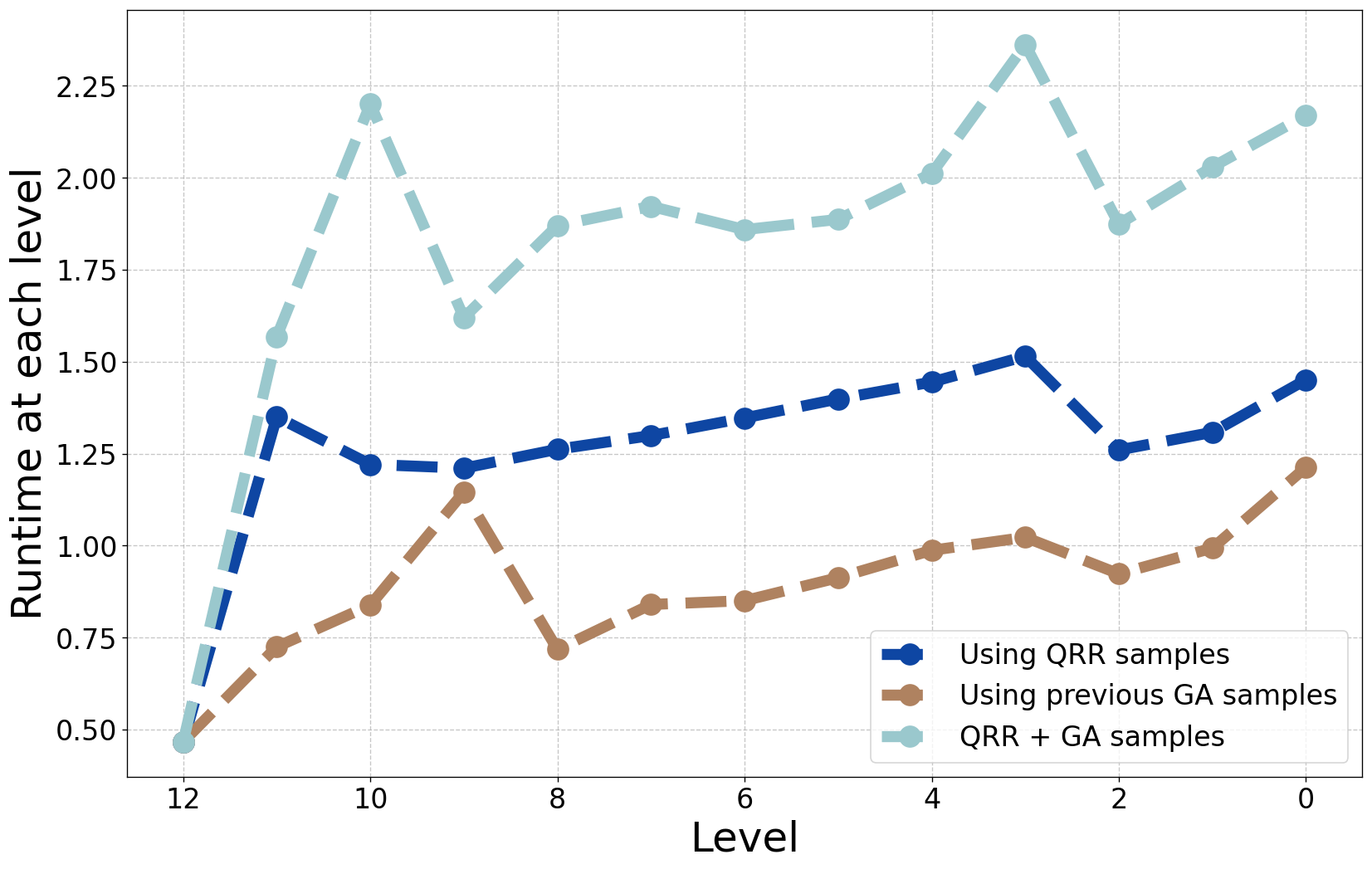}
    \caption{Example of $G_1$ graph from Gset dataset \cite{yyyeGset}. Here, we plot the approximation ratio (top) and the runtime (bottom) for each hierarchy level, where level $12$ denotes the coarsest level, and level $0$ denotes the original graph.}
    \label{fig:QRR+RA_vs_others_approxratio}
\end{figure}

To study the performance of $p=1$ QAOA with QRR within the multilevel framework, we first set up the following experiments with the $G_1$ graph.
We consider three algorithmic variants, where we interpolate solutions between hierarchy levels using (1) only candidate solutions returned by QRR at the previous level; (2) only candidate solutions returned by GA at the previous level, and (3) a combination of both QRR and GA samples from the previous level.
In the scenario (2), the QRR is performed only at the coarsest level to obtain an initial population for the Genetic Algorithm, and all further levels are interpolated using only GA.
In scenarios (1) and (3), we use the QAOA parameter transfer rules outlined previously (see Section~\ref{sec:uncoarsening} and algorithm \ref{algo:Refinement}).
In Figure~\ref{fig:QRR+RA_vs_others_approxratio}, we observe that using only QRR samples to interpolate between levels (scenario (1)) does not improve the coarsest solution, and performs generally worse than using the genetic algorithm initialized with QRR at the coarsest level (scenario (2)).
At the same time, we observe that the combination of both approaches (scenario (3)) yields the best results, demonstrating that the hybrid QRR+GA algorithm is the most performant.

In Table \ref{tab:exp_res}, we present the performance of our proposed method on various graphs, when multilevel solver (either $p=1$ QAOA+QRR+GA (QAOA + BURER02) or Burer-Monteiro (BURER02) is applied to coarsened problems) is used to find an initial-guess solution for non-multilevel Burer-Monteiro SDP implementation.
With the best-found objective, the warm-started Burer-Monteiro is run for $100$ seconds, while getting the initial-guess solution via multilevel algorithm varies between instances, with total runtime of the algorithm being at least $800$s smaller than the non-multilevel Burer-Monteiro implementation to which we compare in the Table.
Here, we compare our results against several non-multilevel classical algorithms, including the rank-2 Semidefinite Programming (SDP) relaxation \cite{burer2002rank}, Greedy Randomized Adaptive Search Procedure \& VNS with path-relinking\cite{festa2002randomized}, a memetic algorithm combined with Variable Neighborhood Search (VNS) \cite{DUARTE2005}, and a tabu search \cite{glover2010diversification}.
We observe that the initialization of the Burer-Monteiro algorithm via solutions obtained from the multilevel solver provides an advantage in the majority of the tested cases, while having reduced total runtime.
At the same time, within the multilevel setting, we observe that our quantum solver outperforms Burer-Monteiro roughly half the time.  
We anticipate that increasing the QAOA depth would further enhance solution quality. 
Overall, these findings highlight the usefulness of our method and how its performance compares to state-of-the-art classical heuristics not based on multilevel strategy.

%% file: sections/6conclusion.tex
\section{Discussion}
We introduced improvements to a highly scalable multilevel QAOA framework that addresses large-scale QUBO problems within the limitations of current NISQ-era hardware. Numerical evidence suggests that finer-level information is carried through coarsened levels and can be employed to forward subsequent refinements, reducing the search space and accelerating convergence. Despite these promising results, several important research directions remain open, such as a) proving the parameter transfer analytically for $p=1$ QAOA+QRR within the multilevel setup for toy problems; b) investigating the performance of our method for higher circuit depth; or c) devising further improvements of the coarsening and refinement schemes, for example via changing the topology and/or dimension of the embedding. 